\makeatletter
\declare@file@substitution{revtex4-1.cls}{revtex4-2.cls}
\makeatother

\documentclass[twocolumn]{aastex631}

\received{January 29, 2025}
\revised{July 21, 2025}
\accepted{August 07, 2025}

\submitjournal{AJ}

\shorttitle{\textsc{NutMaat} for MK Classification}
\shortauthors{R.~I.~El-Kholy \& Z.~M.~Hayman}

\graphicspath{{./}{images/}}

\usepackage{tabularx, xcolor, amssymb}

\begin{document}
	\title{\texttt{\textsc{NutMaat}}: A Python package for stellar spectral classification on the MK system}
	
	\correspondingauthor{R.~I.~El-Kholy}
	\email{relkholy@sci.cu.edu.eg, reham.elkholy@cu.edu.eg}
	
	\author[0000-0001-6337-1384]{R.~I.~El-Kholy}
	\affiliation{Department of Astronomy, Space Science, and Meteorology \\
		Faculty of Science, Cairo University \\
		1 Gamaa Street, Giza 12613, Egypt}
	
	\author[0009-0009-4770-5626]{Z.~M.~Hayman}
	\affiliation{Department of Astronomy, Space Science, and Meteorology \\
		Faculty of Science, Cairo University \\
		1 Gamaa Street, Giza 12613, Egypt}
		
	\begin{abstract}
		Stellar spectral classification according to the Morgan-Keenan (MK) system remains fundamental to astrophysical studies, yet modern surveys require automated, scalable tools. We present \texttt{\textsc{NutMaat}}, an open-source Python-based package inspired by \texttt{MKCLASS}, designed to automate MK classification while addressing scalability and usability limitations. It employs modern computational tools for batch processing and offers a modular architecture that enables efficient, platform-independent analysis of large spectral datasets. It also includes modules for detecting classical chemically peculiar stars, such as Am, Ap, and $\lambda$ Boo types, using internal consistency checks between different line diagnostics. Tested on the CFLIB and MILES libraries, \texttt{\textsc{NutMaat}} achieved spectral and luminosity classification accuracies comparable to \texttt{MKCLASS}, with minimal systematic offsets and a robust performance down to S/N $\lesssim$ 10. \texttt{\textsc{NutMaat}} successfully identified chemically peculiar stars, tested on LAMOST DR7 ACV variables, and processed the SDSS-IV MaStar library---which lacks native MK classifications---to produce a stellar catalog, demonstrating survey readiness. Future development of \texttt{\textsc{NutMaat}} will focus on extending wavelength coverage beyond the 3800--5600 \r{A} range, computational acceleration via Cython, and refining peculiarity classification. Beyond its technical design, \texttt{\textsc{NutMaat}} can provide consistent, MK-standard classification across large-scale spectroscopic surveys, facilitating reliable stellar population analyses, identification of rare objects, and the construction of high-quality spectral catalogs essential for galactic archaeology and stellar evolution studies. As an open-source tool, \texttt{\textsc{NutMaat}} bridges traditional MK methods with modern data workflows, offering a scalable solution for current and future spectroscopic surveys.
	\end{abstract}
	
	
	\section{Introduction}	\label{sec:intro}	
		The Morgan-Keenan (MK) classification system is the main spectral classification system used by astronomers to classify stellar spectra. It was first developed by \citet{Morgan1943}. Stellar spectral types play a crucial role in many astrophysical fields \citep{Gray2009}. In addition to estimation of physical parameters (effective temperature, surface gravity, and metallicity), they provide identification of peculiar stars and those of astrophysical interest. Although the MK system started out in the optical range, it has since been extended into the ultraviolet \citep{Walborn1985,Walborn1995} and the infrared \citep{Andrillat1995,Hanson1996,Meyer1998,Wallace2000} ranges. A thorough explanation of the MK classification system can be found in \citet{Gray2009}.
		
		Spectral classification started out by eye using glass plates. Eventually, digital detectors enabled astronomers to move it to the computer screen. But it was still performed by human inspection of individual spectra. Meanwhile, large astronomical surveys collect increasingly-large amounts of data, emphasizing the need for automatic spectral classification. Thus, automatic classification systems started to develop along two lines; namely, the metric-distance technique \citep{LaSala1994} and the Artificial Neural Networks (ANN) technique \citep{Gulati1994,Singh1998}. Eventually, fuzzy-logic knowledge-based classification systems have been developed \citep{Carricajo2004,Manteiga2009}. The most popular technique among the three has been the ANN one. However, it requires a large set of training spectra to be classified in advance by an expert. Hence, the connection of the classification results with the MK standards are ambiguous \citep{Gray2014}. In contrast, the metric-distance technique is the most faithful to the orthodox way of MK classification, as it still depends on direct comparison with the MK standard spectra; a recent example of this is \texttt{\textsc{PyHammer}} \citep{Kesseli2017,Roulston2020}. While both the metric-distance and the ANN techniques have been successful in temperature classification, the same could not be said of luminosity classification. In addition, the metric-distance method is highly dependent on the accuracy of the flux calibration or the rectification of the involved spectra, while the ANN technique is quite sensitive to the distribution of the training spectra across spectral types. As for the fuzzy-logic technique, it tries to compare spectra to the MK standards using a list of line ratios, line strengths, and band fluxes; as implemented in \texttt{STARMIND} \citep{Manteiga2009}. However, both \texttt{\textsc{PyHammer}} and \texttt{STARMIND} provide one-dimensional spectral types with no luminosity classes. Nevertheless, the fuzzy-logic concept can be usefully integrated within the context of an expert system \citep{Gray2014}.
		
		Convinced of the necessity of a ``simple, adaptable, automatic classification system", \citeauthor{Gray2014} developed the \texttt{MKCLASS}\footnote{\url{http://www.appstate.edu/~grayro/mkclass}} program to classify stellar spectra according to the MK system through direct comparison with the MK standards using procedures that mimic those of human classifiers. Before discussing the capabilities of \texttt{MKCLASS}, we first review how human experts classify spectra. An expert uses a library of standard spectra that had ideally been obtained by the same instrument/telescope combination as the spectra to be classified. If such a library is not available, then the resolutions of the standards and the program spectra are matched by convolving either of them with a carefully chosen line-spread function. A program spectrum is then displayed on the computer screen. The expert starts by assigning a rough spectral type before performing a detailed comparison with the MK standards (usually two standards are displayed for comparison with the program spectrum). Once a refined temperature type is determined by finding two standards of the same preliminary luminosity type that bracket the temperature type of the program spectrum, an improved luminosity type is then found using the same steps along the luminosity axis. This process is iterated until no better type match can be found.
		
		The procedure described so far assumes the spectrum in question to be ``normal". However, there is always a possibility that the spectrum under consideration is ``peculiar" in some way, which consequently means that it will not closely match any of the MK standards. The procedure employed for classification in such cases will completely depend on the type of peculiarity encountered. Because peculiar stars often have great astrophysical interest, human experts take great care in dealing with them. While in some cases it is enough to reach a best spectral type and just append the peculiarity found at the end---such as in K2 III CN2, where ``CN2" indicates that the CN (referring to a diatomic molecule composed of carbon and nitrogen atoms) absorption bands are two steps stronger than normal---in other cases different parts of the spectrum have to be described by different standards due to the dominance of the peculiarity. For example, kA5hF0mF5 II indicates that the calcium K line resembles an A5-type, the hydrogen lines an F0-type, and the metallic-line spectrum an F5-type, while the luminosity class is II. A more thorough discussion of expert classification of spectra can be found in \citet{Gray2009,Gray2014}.
		
		\texttt{MKCLASS} is a ``C" language computer program that only runs on UNIX/Linux Operating Systems (OS). Its function is to try mimicking the human process of spectral classification as closely as possible. It has a \emph{knowledge base}, represented in a library of MK standard spectra, and an \emph{inference engine}, represented by the different modules that use human-like reasoning to carry out the classification process through direct comparison of the program spectrum with the standard spectra. Together, they provide the basic elements of an ``expert system" \citep{Jackson1990}. However, what makes \texttt{MKCLASS} faithful to the original process of the MK classification system is that its knowledge base is embedded in the standard spectra it uses for real-time measurements that are then fed to the inference engine before deciding on a spectral type \citep{Gray2014}. \texttt{MKCLASS} has since proven useful to many studies \citep[e.g.][]{Carbon2018,Tian2021,Huo2024,Kudritzki2024}.
		
		\texttt{\textsc{NutMaat}} is a Python-based package for stellar spectral classification, inspired by \texttt{MKCLASS}\footnote{\texttt{\textsc{NutMaat}} is based on version 1.07 of \texttt{MKCLASS} which was released 2015 September 10.}. It offers a fully Pythonic interface, utilizing \texttt{pandas} \citep{McKinney2010} data frames to enable efficient batch processing of multiple spectra, rather than handling individual files. This shift not only streamlines large-scale spectral analysis but also makes the software OS-independent, ensuring seamless operation across platforms. It aims to simplify spectral classification while enabling scalable and efficient workflows that can be readily integrated in pipelines of astronomical surveys. \texttt{\textsc{NutMaat}} can be installed via \texttt{pip} and its source code is made available on GitHub\footnote{\url{https://github.com/rehamelkholy/NutMaat}} and archived at Zenodo \citep{Elkholy2024}.
		
		The scientific value of stellar spectral classification lies not only in its utility in the estimation of stellar parameters, but also in the critical role it plays in large-scale analyses of stellar populations, galactic structure, and stellar evolution. In the era of large-scale surveys, astronomers are faced with millions of stellar spectra that require rapid, uniform classification. Traditional manual classification is not scalable, and even existing automatic systems often sacrifice fidelity to the MK system for speed or machine learning generalization.
			
		\texttt{\textsc{NutMaat}} provides a practical solution to this problem. By automating MK-classification through a physically-driven, human-mimicking framework, \texttt{\textsc{NutMaat}} ensures close alignment with the established MK standards, which enables the construction of homogeneous catalogs that can be used for:
		\begin{itemize}
			\item the investigation of stellar demographics across different galactic environments;
			\item target selection with specific spectral and luminosity classes for follow-up studies;
			\item tracing star formation history and the Initial Mass Function (IMF);
			\item anomaly detection and peculiar stars identification with rare astrophysical signatures; and
			\item the minimization of systematic errors introduced by inconsistent human or data-driven classification.
		\end{itemize}
		Furthermore, \texttt{\textsc{NutMaat}}'s open-source design and platform independence allow it to be easily integrated to support real-time classification during observations and offer reproducible results. Hence, \texttt{\textsc{NutMaat}} is not only a technical advancement but also a scientific tool for modern astronomical surveys.
		
		In this work, we introduce, describe, and test the capabilities and limitations of \texttt{\textsc{NutMaat}}. The paper is structured as follows: Section \ref{sec:mkclass} provides an overview of \texttt{MKCLASS}, its methodology, and its limitations. Section \ref{sec:nutmaat} introduces \texttt{\textsc{NutMaat}}, describing its design principles. Section \ref{sec:tests} presents testing results using benchmark stellar spectral datasets to evaluate \texttt{\textsc{NutMaat}}'s accuracy and performance. Section \ref{sec:nm-mastar} presents a stellar catalog based on \texttt{\textsc{NutMaat}}'s application to the MaNGA Stellar Library \citep[MaStar;][]{Yan2019} from the seventeenth data release (DR17) of the Sloan Digital Sky Survey \citep[SDSS;][]{Abdurrouf2022}. Finally, Section \ref{sec:discussion} discusses our findings and potential future improvements, before concluding in Section \ref{sec:conclusions}.
	
	\section{\texttt{MKCLASS}}	\label{sec:mkclass}
		In this section, we give a review of the most important features, capabilities, and limitations of \texttt{MKCLASS} while a more thorough discussion of the software can be found in \citet{Gray2014} and the accompanying documentation \citep{Gray2014a}. Two libraries of MK standard spectra are provided to \texttt{MKCLASS}. The first is \texttt{libr18} with wavelengths ranging from 3800--4600 \r{A} contains 1.8 \r{A} resolution rectified spectra. The second is \texttt{libnor36} with wavelengths ranging from 3800--5600 \r{A} contains 3.6 \r{A} resolution flux-calibrated and normalized spectra. All fluxes were corrected, dereddened, and normalized at 4503 \r{A}. In addition, it is possible to add custom spectral libraries to \texttt{MKCLASS}. The classification criteria of the program covers the violet-green part of the spectrum (3800--5600 \r{A}). However, the essential criteria fall in the range 3918--4600 \r{A} for O- to K-stars and extend to 5000 \r{A} to include M-stars. \texttt{MKCLASS} requires a subgrid fully populated with MK standards of predetermined temperature types and at luminosity classes ``V," ``III," ``Ib," and ``Ia." The subgrid does not have to span the entire spectral type range, but a subrange can be specified in a configuration file. If MK standards are not available at all grid points, interpolation can be applied between existing MK standards while building the library.
		
		\texttt{MKCLASS} includes a companion program, \texttt{MKPRELIM}, that carries out the preliminary processing of the program spectra. It determines the radial velocity of the spectrum and transforms it to the star's rest frame. If the spectrum is in the flux or uncalibrated format, it will also normalize it at a specific wavelength. If the resolution of the spectrum does not match that of the spectral library, other auxiliary modules are included in the program that can be used to convolve the spectrum with a line-spread function.
		
		\texttt{MKCLASS} begins the classification process by estimating a rough spectral type, using either the \texttt{ROUGHTYPE1} or \texttt{ROUGHTYPE2} module depending on the flux calibration status of the input spectrum. The \texttt{ROUGHTYPE1} module employs a weighted $\chi^2$ difference between the input and standard spectra and is faster, making it suitable for approximately calibrated or highly noisy data. In contrast, \texttt{ROUGHTYPE2} uses spectral indices---numerical measurements such as line depths, band strengths, or flux ratios over diagnostic wavelength intervals---to assess key classification features. Although this method is computationally slower, it is generally recommended due to its higher robustness against normalization artifacts and its closer adherence to MK classification criteria. This step also identifies whether the input spectrum is broadly consistent with MK standards or exhibits characteristics that fall outside the domain of the MK system---such as strong emission lines, peculiar continua, or noisy, incomplete coverage. If the spectrum deviates significantly from MK morphology, it is flagged as a non-MK type, and the program outputs a coarse classification without proceeding to detailed interpolation or $\chi^2$ analysis. This mechanism ensures that computational resources are not expended on spectra that are unsuitable for MK-based classification. Once the program has made sure that the spectrum is not a non-MK type and a rough type has been determined, the program spectrum undergoes a process of detailed comparison with the standard library for which \texttt{MKCLASS} uses five modules. These modules are responsible for classifying O-stars, B-stars, A-stars, F- and G-stars, and K- and M-stars. The comparison process is carried out by subroutines invoked by those five modules; and if at any point a subroutine indicates that a star belongs to a different type, the spectrum is passed to the corresponding adjacent module.
		
		Mimicking a human classifier, \texttt{MKCLASS} checks for spectral peculiarities at every step of the classification process---a design choice that distinguishes it from other automatic classification systems and necessitates a more sophisticated approach than simple least-square fitting to spectral libraries. As described by \citeauthor{Gray2014}, the program adopts different strategies for different spectral classes. In A-type stars, for instance, \texttt{MKCLASS} independently determines temperature types based on the Ca$\,$\textsc{ii} K line, the Balmer lines, and the general metallic-line spectrum. A spectrum is assumed to be normal only if these independently derived types are in reasonable agreement. If the Ca$\,$\textsc{ii} K line and metallic-line types are earlier than the hydrogen-line type, \texttt{MKCLASS} suspects a metal-weak star, such as $\lambda$ Boo star or a horizontal-branch object. If the K-line type is earlier and the metallic-line type later than the hydrogen-line type, the program identifies the spectrum as an Am (metallic-line A-type) star and passes it to a specialized classification routine. Additionally, \texttt{MKCLASS} checks for chemically peculiar Ap stars by examining specific indicators such as Sr$\,$\textsc{ii} $\lambda4077$ and Si$\,$\textsc{ii} $\lambda\lambda4128-30$. The spectral classification criteria employed by \texttt{MKCLASS} differ systematically across the Hertzsprung-Russell (HR) diagram, with distinct diagnostic features emphasized for each spectral and luminosity class. A detailed description of these criteria can be found in \citet{Gray2009}. Since any necessary use of line ratios or spectral indices is carried out during the classification process, adding custom standard stellar libraries to \texttt{MKCLASS} is fairly easy which makes it adaptable to large spectral surveys and a wide range of spectral resolutions.
		
		\texttt{MKCLASS} does not require dereddening of program spectra and is fairly tolerant of low signal-to-noise (S/N) ratios. This is because \texttt{MKCLASS} only uses line ratios for lines with small wavelength separations. Otherwise, criteria involving spectral feature strengths are measured using spectral indices by centering the feature in question in a narrow band and measuring the line flux before measuring the ratio with fluxes in ``continuum" bands that flank the feature. In addition to line ratios and spectral indices, \texttt{MKCLASS} also utilizes direct comparison of several large spectral regions depending on the spectral type. However, this does not necessitate accurate rectification or flux calibration as approximate normalization is applied to the regions in question prior to the comparison using a low-order polynomial fit.
		
		As it converges to a solution, \texttt{MKCLASS} interpolates between standard spectra, iterating until a precise spectral type is reached. A user can adjust the number of iterations such that \texttt{MKCLASS} repeats the classification process starting with the final spectral type determined during the previous round. This improves the classification precision by starting the process at a better initial point. This repetition can also be exploited to improve flux calibration and the precision of the final type for spectra with poor initial flux calibration. It is also possible to enable an option that outputs a flux corrected version of the spectrum, which may be useful in estimating basic stellar parameters.
		
		The classification results are saved to two files. The first, called the ``classification file", includes only the final spectral type and a quality evaluation---if the star is determined to be ``normal"---based on an overall $\chi^2$ comparison with the best match from the standard library. The second file is a ``log" containing detailed information on the steps employed by the software to reach the final spectral type, which is useful for both debugging purposes and giving the user insight into the nature of peculiar stars. In addition, for normal stars, a spectrum is written to an \texttt{ASCII} file based on the interpolation of the best match from the standard library. Spectra can be batched for classification as \texttt{MKCLASS} is operated from the command line and results are appended to the two files mentioned above. However, the software can only operate on a UNIX/Linux OS. In addition, each spectrum has to be saved in a separate \texttt{ASCII} file with two columns: the first is the wavelength in angstroms and the second is the flux or rectified intensity. Thus, for classification of large batches of spectra, \texttt{MKCLASS} requires quite some preparation before its application to the batch.
	
	\section{\texttt{\textsc{NutMaat}}}	\label{sec:nutmaat}
		The \texttt{\textsc{NutMaat}} Python package, available on GitHub and via PyPi\footnote{\url{https://PyPi.org/project/NutMaat}}, is designed as an equivalent to \texttt{MKCLASS}. It utilizes the same methodologies described in \citet{Gray2014} to mimic human classification of stellar spectra while employing Python's ecosystem to enhance usability, scalability, and integration with modern data analysis workflows and pipelines. \texttt{\textsc{NutMaat}} includes the same MK standard libraries available in \texttt{MKCLASS} with the possibility to add custom libraries as well. It uses the same preprocessing and classification scheme but employs the object-oriented framework available to make the software modular and sustainable for future development. Moreover, both systems include procedures for identifying and labeling chemically peculiar stars when they deviate from the expected patterns.
			
		Despite their conceptual similarity, the two tools differ in significant technical respects. \texttt{MKCLASS} is written in C and operates only on UNIX/Linux systems, requiring users to prepare input spectra as separate \texttt{ASCII} files. It is operated from the command line, and although it allows batch processing through executable batch files, the process is not optimized for large-scale automated workflows. In contrast, \texttt{\textsc{NutMaat}} is fully written in Python and is compatible with all major operating systems. It supports the processing of large batches of spectra in the form of \texttt{pandas} \citep{pandas2024, McKinney2010} data frames, which streamlines classification in survey-scale contexts. Similarly, if the user chooses so, \texttt{\textsc{NutMaat}} can return the classification results as a \texttt{pandas} data frame with columns including spectral and luminosity types, match quality (a categorical string indicating fit confidence: `Excel', `Vgood', `Good', `Fair', or `Poor' based on $\chi^2$ comparison with MK standards and the S/N of the input spectrum), and a peculiarity flag (a string denoting specific chemical/line anomalies---e.g., `Sr', `Eu', or `Cr'---if detected; empty otherwise). Crucially, regardless of whether this \texttt{pandas} data frame is requested or not, \texttt{\textsc{NutMaat}} will always generate the two standard result files described in Section \ref{sec:mkclass}. In addition, it is still possible to both choose the number of iterations to enhance the classification precision and output an adjusted version of the spectrum (where flux corrections have been applied during the preprocessing stage) as an \texttt{ASCII} file for use as input in subsequent classification runs to further refine results.
			
		Classification results from \texttt{\textsc{NutMaat}} and \texttt{MKCLASS} are broadly consistent. Minor differences in output arise due to the use of distinct numerical libraries: \texttt{\textsc{NutMaat}} employs routines from the \texttt{SciPy} \citep{Virtanen2020} package for interpolation and optimization, while \texttt{MKCLASS} relies on custom C-based implementations. Nevertheless, statistical comparisons presented in Section \ref{subsec:sptlum} show that these differences are negligible in practice, with both systems achieving comparable classification accuracy.
			
		In addition to \texttt{MKCLASS}, other automatic classification systems have been developed that prioritize usability and accessibility over strict adherence to MK classification criteria. The most prominent among these are \texttt{\textsc{PyHammer}} and \texttt{STARMIND}. \texttt{\textsc{PyHammer}} \citep{Kesseli2017, Roulston2020} is a Python-based tool designed to classify stellar spectra by computing $\chi^2$ differences between input spectra and a set of empirical templates, and it returns the best-matching spectral type. While \texttt{\textsc{PyHammer}} is simple to use and integrates well into Python-based workflows, it has several limitations. It does not provide luminosity classifications, nor does it attempt to identify chemically peculiar stars. \texttt{STARMIND} \citep{Manteiga2009}, by contrast, uses a fuzzy-logic approach to perform classification based on predefined line ratios, line strengths, and flux bandpasses. This methodology provides some degree of interpretability and robustness to noise. However, \texttt{STARMIND} does not output luminosity classes, and like \texttt{\textsc{PyHammer}}, it does not incorporate any explicit mechanism for detecting or labeling spectral peculiarities. Moreover, the software is implemented in OPS-R2 \citep{Forgy1981} and is not publicly available in open-source form, which limits its accessibility and reproducibility.
			
		\texttt{\textsc{NutMaat}} offers a distinct combination of features that sets it apart from these alternatives. Like \texttt{\textsc{PyHammer}}, it is Python-based, but it also provides MK-consistent luminosity classes and includes a decision-tree structure that enables identification of peculiar stars when they deviate from expected MK standards. Unlike \texttt{STARMIND}, which employs a static set of spectral indicators within a fuzzy-logic framework, \texttt{\textsc{NutMaat}} implements dynamic, type-specific decision trees that emulate human judgment in spectral comparison. Furthermore, unlike \texttt{\textsc{NutMaat}}, neither \texttt{\textsc{PyHammer}} nor \texttt{STARMIND} supports batch processing through data frames. These differences make \texttt{\textsc{NutMaat}} uniquely suited for modern spectroscopic surveys. A comparative overview of these tools is provided in Table$\,$\ref{tab:comparison}, summarizing their accessibility, features, and limitations.
		
		\begin{table*}
			\caption{Feature comparison for the different stellar classification tools mentioned above.}
			\label{tab:comparison}
			\setlength{\tabcolsep}{7pt}
			\centering
			\begin{tabular}{lllccc}
				\hline\hline
				Tool						& Platform 				& Language	& Luminosity Class	& Data Frame Batching	& Peculiarity Detection\\
				\hline
				\texttt{STARMIND}    		& UNIX/Linux (Legacy)	& OPS-R2 	& \texttimes 		& \texttimes 			& \texttimes\\
				\texttt{\textsc{PyHammer}}	& Cross-platform		& Python 	& \texttimes 		& \texttimes 			& \texttimes\\
				\texttt{MKCLASS}			& UNIX/Linux only		& C 		& \checkmark		& \texttimes 			& \checkmark\\
				\texttt{\textsc{NutMaat}}  	& Cross-platform		& Python 	& \checkmark 		& \checkmark 			& \checkmark\\
				\hline
			\end{tabular}
		\end{table*}

		While Section \ref{sec:tests} shows that \texttt{\textsc{NutMaat}} performs comparably to \texttt{MKCLASS}, a rigorous, tool-agnostic benchmark testing set is essential for formal validation. To this end, we are planning to curate a dataset that will enable direct comparisons among \texttt{\textsc{NutMaat}}, \texttt{MKCLASS}, \texttt{\textsc{PyHammer}}, and \texttt{STARMIND} (where feasible). It will include 1000--2000 spectra drawn from several benchmarking libraries, selected to span a broad range of spectral types, luminosity classes, and S/N. Expert-verified classification from the literature will serve as reference labels. While no public-facing user guide is currently available, a full documentation set---including a Wiki page on the official GitHub repository---is in preparation and will accompany the next release.

	\section{Testing \texttt{\textsc{NutMaat}}}	\label{sec:tests}
		\subsection{Spectral and Luminosity Classes} \label{subsec:sptlum}
			When \citeauthor{Gray2014} first tested \texttt{MKCLASS}, they curated a set of stellar spectra taken mostly from the northern sample \citep{Gray2003} of the NStars Project, in addition to a supplementary set of A-, \mbox{F-,} and G-type stars \citep{Gray2001}. However, the two spectral standard libraries included with \texttt{MKCLASS} are obtained from spectra taken by the same spectrograph/telescope combination as these test sets. In addition, the classification of these sets was carried out by the developers of \texttt{MKCLASS}. Hence, the test results obtained by \citeauthor{Gray2014} are the best results that may be expected from \texttt{MKCLASS}. They found that the software classification, in comparison with the expert human classification, had a standard deviation of 0.59 spectral subclass with a systematic difference of $0.08 \pm 0.02$ subclass. For luminosity classification, the standard deviation was 0.52 luminosity class with a systematic difference of only $0.02 \pm 0.03$ luminosity class.
			
			To test \texttt{\textsc{NutMaat}}'s ability to accurately classify stellar spectra, we have used two widely used empirical stellar libraries: the Indo-U.S. Library of Coud\'{e} Feed Stellar Spectra \citep[CFLIB;][]{Valdes2004} and the Medium-resolution Isaac Newton Telescope Library of Empirical Spectra \citep[MILES;][]{SanchezBlazquez2006}. The CFLIB, also known as Indo-US library, includes spectra of 1273 stars that were captured at a resolution of $\sim 1$ \r{A} (R $\sim$ 5000) full width at half-maximum (FWHM) over the wavelength range 3460--9464 \r{A} and they cover a broad range of parameter space in terms of effective temperature, surface gravity, and iron metallicity. Observations were taken of 6917 individual spectra for 1273 stars over a period of eight years using five different grism settings to cover the entire desired wavelength range. The MILES library contains 985 flux-calibrated stellar spectra at a FWHM resolution of 2.56 \r{A} (R $\sim$ 2000) covering the wavelength range 3536--7410 \r{A} and normalized at 5550 \r{A}. Like CFLIB, MILES also spans a wide range of stellar parameters and was constructed with the goal of providing uniform spectral calibration across the HR diagram.
			
\begin{figure*}
				\centering
				\includegraphics[width=\textwidth]{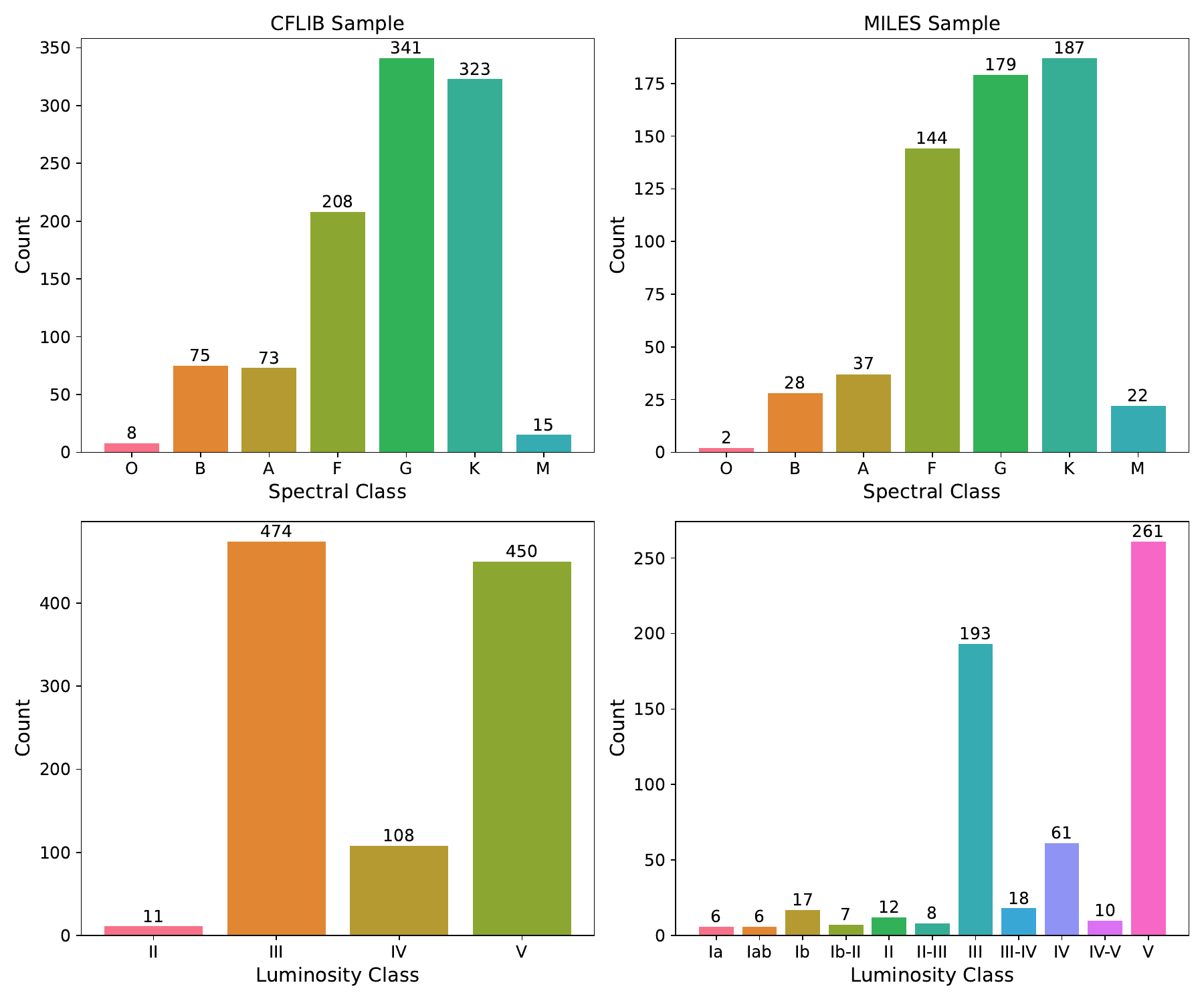}
				\caption{Distribution over main spectral and luminosity classes for the stars from the CFLIB and MILES data bases. The distribution is shown only for those stars which are selected for this study and not for all the spectra in these libraries.}
				\label{fig:test-set-distro}
			\end{figure*}

			For the CFLIB spectra, the classification of each spectrum was extracted from the header of the \texttt{fits} file\footnote{\url{https://noirlab.edu/science/observing-noirlab/observing-kitt-peak/telescope-and-instrument-documentation/cflib}}; while the classification of the MILES spectra was obtained from their associated VizieR catalog \citep{SanchezBlazquezCat}. In addition, the CFLIB spectra were convolved with a Gaussian to reduce the resolution to that of the \texttt{libnor36} MK standards library. After dropping instances that were unclassifiable by one or both of \texttt{MKCLASS} and \texttt{\textsc{NutMaat}}, or where the original classification was invalid, we ended up with 1043 spectra from CFLIB and 599 spectra from MILES. The distribution of spectra used in the final test set over temperature and luminosity classes are shown in Figure \ref{fig:test-set-distro}.
			
\begin{figure}
				\centering
				\includegraphics[width=\columnwidth]{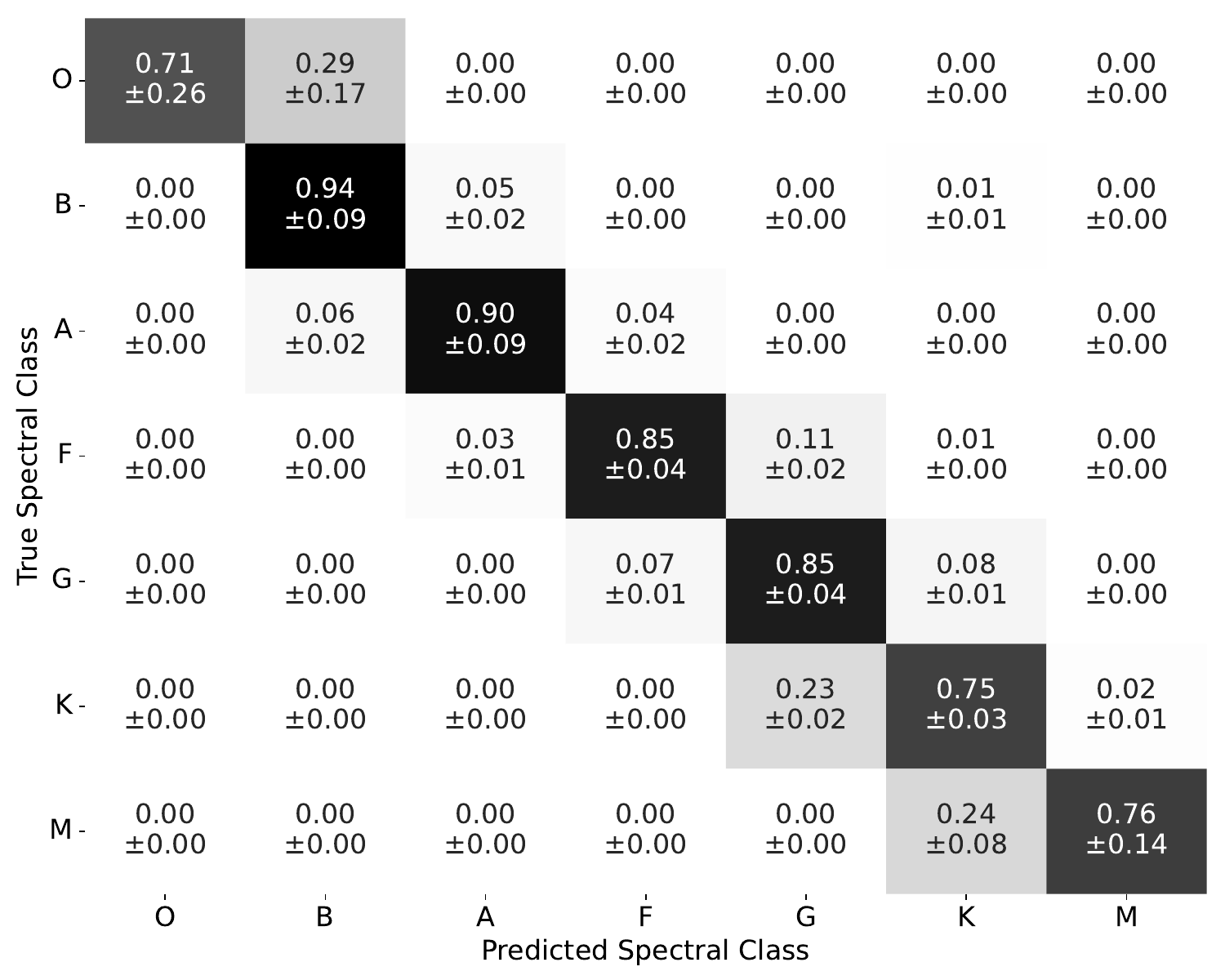}
				\caption{The confusion matrix of the main temperature type classification using \texttt{\textsc{NutMaat}} for the merged CFLIB and MILES test sets.}
				\label{fig:confusion-matrix}
			\end{figure}

			Figure \ref{fig:confusion-matrix} shows the confusion matrix of the main temperature type classification using \texttt{\textsc{NutMaat}} for the test set. While it demonstrates that \texttt{\textsc{NutMaat}} performs well across most spectral types, the statistical reliability of individual matrix entries---particularly for underrepresented types such as O and M stars---requires careful interpretation. In the test set, the number of O-type spectra is especially small (e.g., 8 in CFLIB and 2 in MILES; see Figure \ref{fig:test-set-distro}), meaning that even one or two misclassifications (e.g., an O9 as B0) can cause large fluctuations in confusion matrix entries. To assess the stability of these results, we applied bootstrap resampling \citep{Efron1979}---a statistical technique in which multiple synthetic datasets are created by sampling with replacement from the original data. Over 1000 bootstrap iterations, we computed a new confusion matrix each time and then averaged the results to obtain the mean and standard deviations for each matrix cell. As shown in Figure \ref{fig:confusion-matrix}, the classification uncertainty---reflected in the standard deviation---is significantly larger for sparsely populated classes, quantifying the inherent variability introduced by low-number statistics.
				
			Several of the observed misclassifications in Figure \ref{fig:confusion-matrix} occur between neighboring spectral types---for instance, M to K or K to G---rather than between distant types. These confusions can often be attributed to underlying physical similarities in their spectra within the 3800--5600 \r{A} window used for classification. For late-type stars, the temperature-sensitive molecular bands (e.g., TiO and VO) that define M-type spectra become prominent only toward longer wavelengths, limiting the discriminative power in the \texttt{\textsc{NutMaat}} spectral range. This can lead to M-type spectra being classified as late K-types, especially under lower S/N conditions. Similarly, K stars may be misclassified as G types because the metal-line complexes that differentiate them (e.g., the Mg$\,${\sc i} triplet, Fe$\,${\sc i} blends, and Ca$\,${\sc i} features) evolve gradually with temperature and can be masked at moderate resolution or in spectra affected by line broadening or flux calibration errors. At the early-type end, O stars may be misclassified as B types due to the weakness of He$\,${\sc ii} lines at shorter wavelengths and the dominance of Balmer and He$\,${\sc i} lines, which are also strong in early B-type spectra. The limited number of helium or ionized metal lines in this range makes precise subclass discrimination difficult without broader spectral coverage. These classification ambiguities reflect known degeneracies in the MK system when constrained to moderate-resolution spectra and a limited wavelength window \citep{Gray2009}, which supports the conclusion that they arise from astrophysical and observational constraints, rather than from algorithmic failure in the classification process. Future versions of \texttt{\textsc{NutMaat}} will extend the classification region beyond 5600 \r{A} to include additional helium lines, molecular bands, and low-gravity indicators, which are expected to reduce these degeneracies and improve classification robustness.
			
\begin{figure}
				\centering
				\includegraphics[width=\columnwidth]{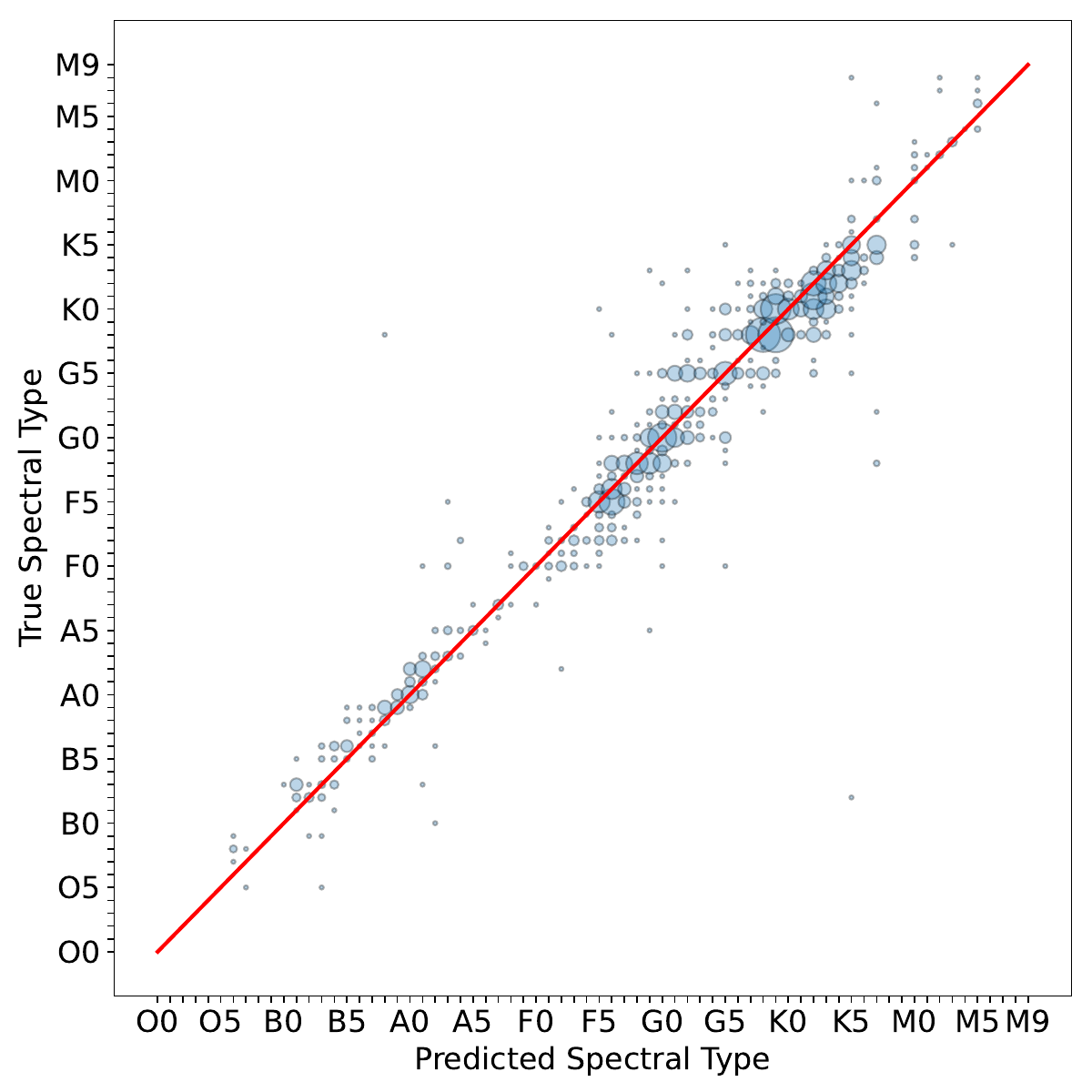}
				\caption{\texttt{\textsc{NutMaat}} temperature types for the CFLIB and MILES sets plotted against the published temperature types of \citet{Valdes2004} and \citet{SanchezBlazquez2006}. The size of a circle is proportional to the number of spectra represented by it. A total of 1642 spectra were used to construct this figure.}
				\label{fig:predicted-true-spt}
			\end{figure}

			Figure \ref{fig:predicted-true-spt} shows the comparison between the \texttt{\textsc{NutMaat}} spectral (temperature) types and the published spectral types from the CFLIB and MILES libraries, where the size of a circle is proportional to the number of spectra represented by it and the solid line has a slope of unity and an intercept of zero. The standard deviation in the comparison is 2.79 spectral subclasses with a systematic difference between the \texttt{\textsc{NutMaat}} and human types of $-0.25 \pm 0.07$ subclass; that is, the \texttt{\textsc{NutMaat}} types are slightly later than the human types. Applying \texttt{MKCLASS} to  the same test set, we obtained a standard deviation of 2.73 subclasses with a systematic difference of $-0.20 \pm 0.07$ subclass.
			
			Figure \ref{fig:predicted-true-spt} also reveals a small number of extreme outliers, such as spectra with true types of K misclassified as B-type, or vice versa. An initial inspection of these cases indicates that they are caused by divergence or instability in the numerical optimization algorithm used during the initial type estimation stage. This typically results in poor convergence or spurious minima in the presence of weak or noisy features. Additional contributing factors may include (1) intrinsic peculiarities in the stellar spectrum, such as unusual chemical abundances or composite systems not yet modeled by \texttt{\textsc{NutMaat}}, and (2) morphological degeneracies where, for instance, the weakened metal-line spectrum of a metal-poor K-type star can resemble that of an early-type object over a restricted wavelength range. These cases are generally associated with low match-quality scores in the output, allowing users to filter or investigate them further, and are rare relative to the overall classification set. As more such instances are accumulated and analyzed in future testing, their root causes can be better isolated and addressed in subsequent versions of \texttt{\textsc{NutMaat}}.
			
\begin{figure}
				\centering
				\includegraphics[width=\columnwidth]{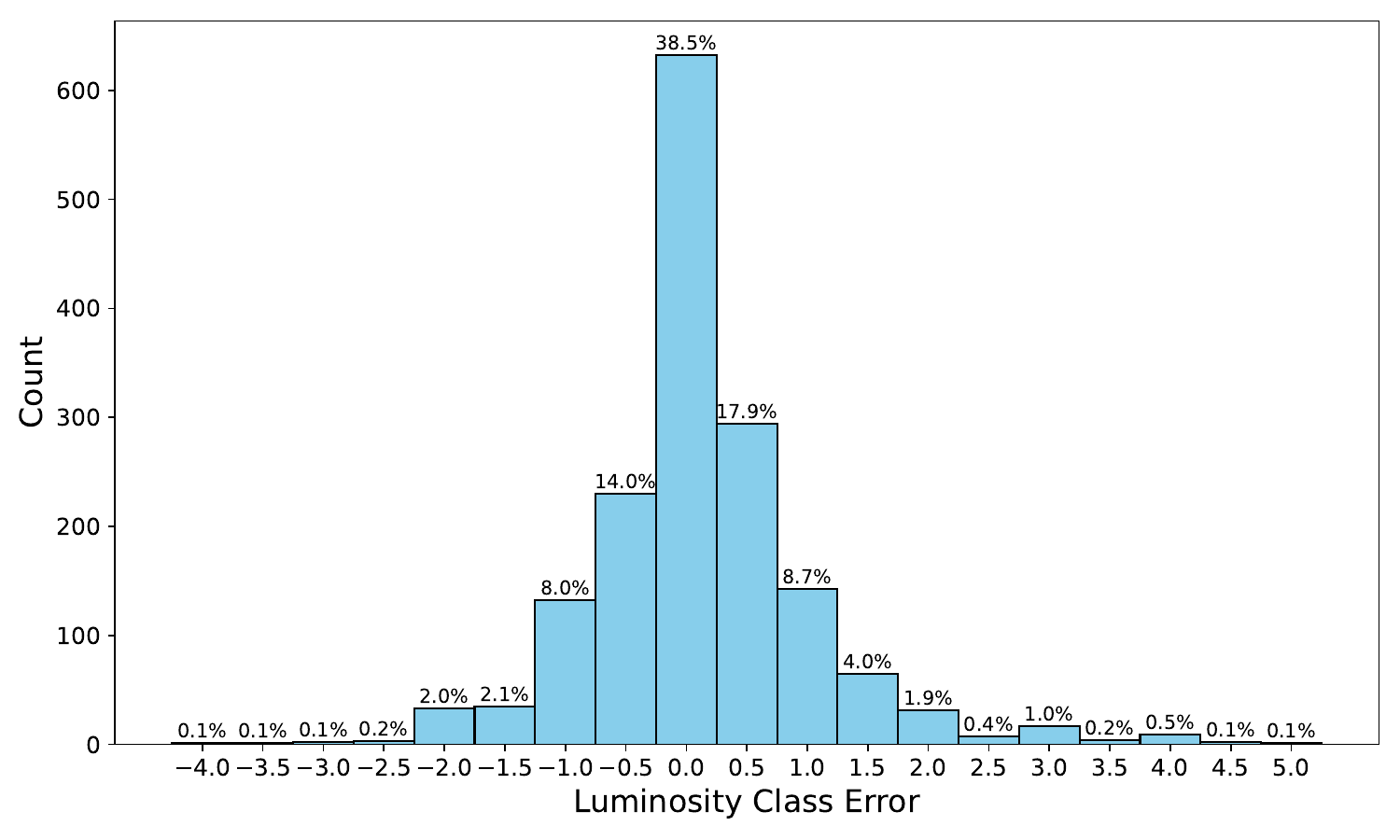}
				\caption{Histogram showing the difference in luminosity class between \texttt{\textsc{NutMaat}} and the human types, where a difference between a luminosity class of II and III, for example, corresponds to a difference of one.}
				\label{fig:lum-error-hist}
			\end{figure}

\begin{figure}
				\centering
				\includegraphics[width=\columnwidth]{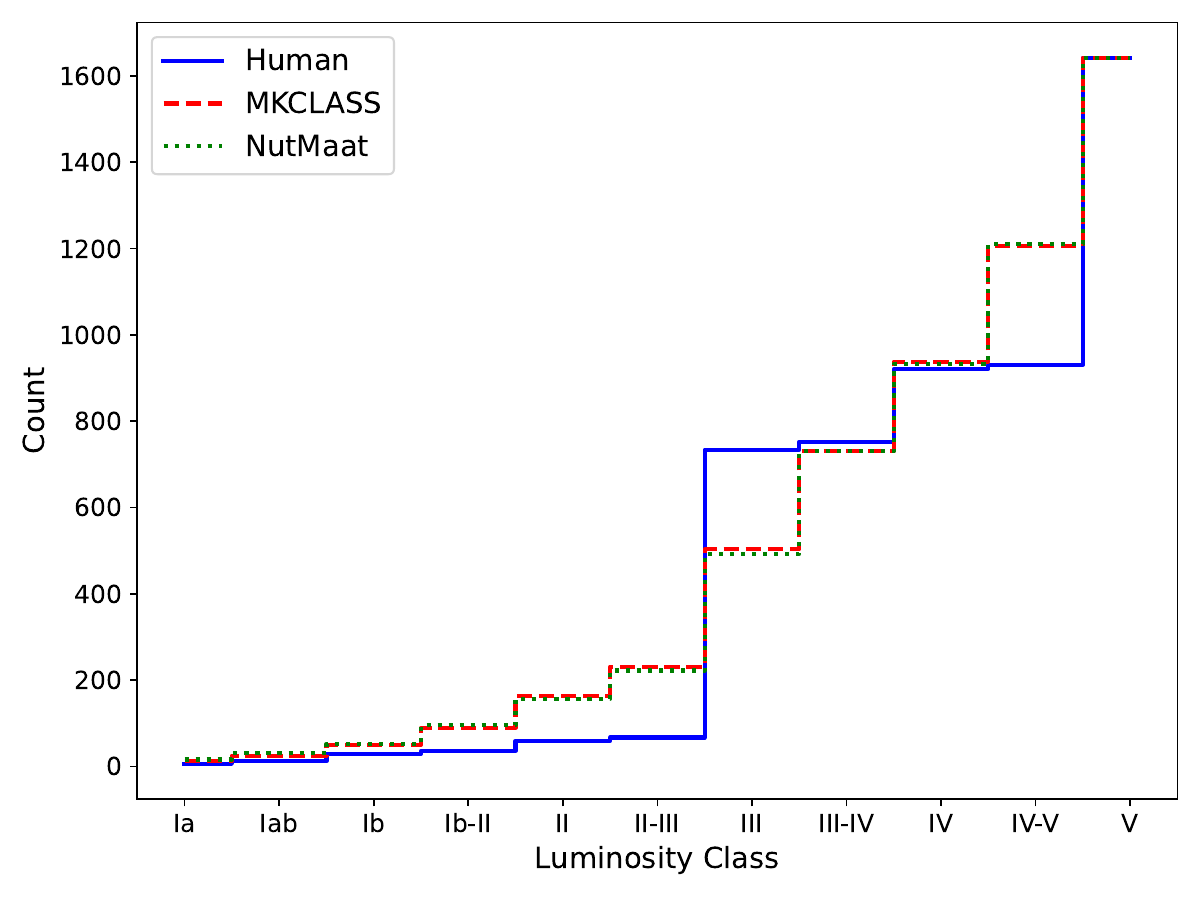}
				\caption{Cumulative luminosity-class distributions for the human, \texttt{MKCLASS}, and \texttt{\textsc{NutMaat}} types. The number of stars with luminosity classes more luminous than the abscissa is plotted to form the cumulative distribution.}
				\label{fig:lum-cumulative}
			\end{figure}

			Figure \ref{fig:lum-error-hist} shows a histogram of the difference in luminosity class between \texttt{\textsc{NutMaat}} types and the human types. The difference has a standard deviation of 0.92 luminosity class, with a systematic difference of only $0.12 \pm 0.02$ luminosity class. Applying \texttt{MKCLASS} to the same test set produced a standard deviation of 0.91 luminosity class, with a systematic difference of $0.12 \pm 0.02$ luminosity class. Figure \ref{fig:lum-cumulative} shows the cumulative luminosity-class distributions for the \texttt{\textsc{NutMaat}}, the \texttt{MKCLASS}, and human types. We can see that both \texttt{\textsc{NutMaat}} and \texttt{MKCLASS} have a tendency to classify dwarfs (low-luminosity stars) on average as slightly more luminous, and giants as slightly less luminous.
			
			To further assess the statistical and astrophysical impact of \texttt{\textsc{NutMaat}}'s luminosity classification, we analyzed the distribution of absolute offsets shown in Figure \ref{fig:lum-error-hist}. We find that 70.4\% of classifications fall within $\pm0.5$ luminosity classes and 87.1\% within $\pm1.0$. However, 12.9\% of the stars differ by more than one full luminosity class, potentially indicating substantial misclassifications (e.g., giants misidentified as main-sequence stars or vice versa). Such deviations can distort inferred gravity distributions or skew the relative frequencies of stellar populations across luminosity classes. These results call for further interpretive tools and error diagnostics. However, these histogram and cumulative offset distributions currently offer a more robust alternative to more advanced analysis tools (e.g., confusion matrices) for evaluating luminosity class performance, especially given the limitations in the underlying reference data. The granularity of luminosity labels varies significantly across spectral libraries: the CFLIB dataset typically includes only broad classifications (e.g., III or V), whereas MILES provides more nuanced subclasses (e.g., IV-V). This inconsistency, along with the unequal representation of luminosity classes (see Figure \ref{fig:test-set-distro}), makes it impractical to apply a bootstrapped confusion matrix analysis as was done for spectral types. Any such approach would require artificial coarsening or interpolation of labels, thereby undermining statistical validity. For this reason, we rely on the offset histogram (Figure \ref{fig:lum-error-hist}) and cumulative error profiles (Figure \ref{fig:lum-cumulative}) to assess performance. A more rigorous statistical treatment of luminosity classification will be possible in future work based on uniformly labeled datasets, as outlined in our benchmarking plan (see Section \ref{sec:nutmaat}). We will then be able to isolate the sources of large-offset cases, characterize their impact, and improve \texttt{\textsc{NutMaat}}'s calibration accordingly.
				
			While the classification logic in \texttt{\textsc{NutMaat}} is explicitly designed to reproduce the decision pathways of \texttt{MKCLASS}, minor deviations occasionally appear in the assigned spectral types and luminosity classes. These differences are not the result of fundamental discrepancies in the classification criteria, but rather stem from the distinct numerical routines used in the two implementations. \texttt{MKCLASS}, written in C, relies on hand-optimized interpolation, fitting, and optimization algorithms tailored for computational efficiency and stability. In contrast, \texttt{\textsc{NutMaat}} employs standard Python libraries such as \texttt{scipy.interpolate} and \texttt{scipy.optimize.minimize}, which, while robust and widely validated, may behave differently in edge cases---for example, in convergence tolerance, local minima handling, or spline smoothness. However, due to the limited number of cases exhibiting discrepancies in our current tests, we refrain from making any general statistical claims at this stage. A more detailed error analysis will require a significantly larger sample of such instances in order to characterize their frequency, directionality, and astrophysical significance. This effort will form part of the broader benchmarking initiative to compare \texttt{\textsc{NutMaat}} against \texttt{MKCLASS} and other classification tools using a controlled, cross-validated dataset. Nevertheless, these results demonstrate \texttt{\textsc{NutMaat}}'s capability of classifying normal stars. On a Dell Inspiron 5567 (Intel Core i7-7500U @ 2.70GHz processor, 8GB DDR4 RAM, Windows 10 Pro 64-bit), \texttt{\textsc{NutMaat}} takes 7--9 seconds to classify a spectrum using three iterations, with overhead preprocessing (data I/O, wavelength selection, normalization/resolution matching, radial velocity correction) consuming the majority of this time.
			
		\subsection{Peculiar Stars Classification} \label{subsec:peculiar}
			As described in Section \ref{sec:mkclass}, \texttt{MKCLASS}---and by extension \texttt{\textsc{NutMaat}}---accounts for spectral peculiarities by evaluating the internal consistency between different temperature diagnostics (e.g., the Ca$\,$\textsc{ii} K line, hydrogen lines, and metallic-line spectrum), and by testing for the presence of the anomalous spectral features such as Sr$\,$\textsc{ii} $\lambda4077$ or Si$\,$\textsc{ii} $\lambda\lambda4128-30$ \citep{Gray2014}. This logic is key to reproducing human-like classification decisions for chemically peculiar stars.
				
			In the MK system, peculiar stars are those whose spectra deviate from the morphology expected of a star with normal chemical composition at a given temperature and luminosity. These deviations may reflect over- or under-abundances of specific elements (e.g., Sr, Eu, Si, Cr, Fe), the presence of molecular bands like CN or CH, or spectral components with different physical origins \citep{Gray2009}. Am stars, for example, typically show enhanced metal-line absorption and are classified by inconsistent line types (e.g., kA1hA5mF2), while Ap stars show strong rare-earth (lanthanides such as Eu, Nd, and Ce) or silicon lines \citep{Preston1974,Smith1996}. $\lambda$ Boo stars exhibit weakened metallic lines compared to their hydrogen-line types \citep{Faraggiana1998,Jura2015}.
				
			Peculiar stars are astrophysically significant because they offer insight into processes not observable in normal stars. Chemically peculiar stars, especially of the Ap and Am types, provide empirical evidence for radiative diffusion, magnetic field structures, and rotational modulation of atmospheric composition. $\lambda$ Boo stars are thought to be linked to selective accretion in early-type systems, while CN- or CH-enhanced giants offer clues about nuclear processing and dredge-up. Identifying such stars is critical not only for accurate classification but also for probing the physical mechanisms governing stellar atmospheres and evolution.
				
			$\alpha^2$ Canum Venaticorum (ACV) variables are a subclass of chemically peculiar stars, typically belonging to the Ap spectral class. These stars are characterized by strong, stable magnetic fields and surface abundance inhomogeneities that lead to overabundances of elements such as strontium (Sr), europium (Eu), and chromium (Cr). As the star rotates, the observed line strengths of these elements vary due to the uneven surface distributions, producing characteristic spectroscopic variability \citep{Samus’2017}. These stars are especially valuable for testing the robustness of automated detection of chemical peculiarities due to their variability. To evaluate \texttt{\textsc{NutMaat}}'s capability in this domain, we applied it to a curated sample of 15 ACV variables from the Large Sky Area Multi-Object Fiber Spectroscopic Telescope \citep[LAMOST;][]{zhao2012} seventh data release (DR7)\footnote{\url{https://dr7.lamost.org/}}, cross-matched with the catalog of chemically peculiar stars by \citet{Faltova2021}. The LAMOST low-resolution spectra cover the wavelength range 3690--9100 \r{A}, with a resolution of R $\sim$ 1800. These stars were not selected because of their variability, but because they constitute a consistently classified and uniformly observed set of chemically peculiar stars---essential for isolating classification performance without instrumental or methodological differences. While ACV stars are known to be both photometrically and spectroscopically variable, this is not the focus of our analysis. \texttt{\textsc{NutMaat}} does not explicitly model such phase-dependent variation, but its classification logic is designed to identify spectral anomalies robustly even in single-epoch observations.
				
			\texttt{\textsc{NutMaat}} correctly flagged most of the expected peculiarities, consistently identifying the presence of Sr, Si, and Eu lines, even in cases where the temperature or luminosity class differed from literature classifications. Table \ref{tab:peculiars} presents these results. While the MK types produced by \texttt{\textsc{NutMaat}} may not exactly match the human-assigned classifications, the consistent flagging of anomalies (in 14 out of 15 instances) indicates that its peculiarity detection logic is functioning effectively.
			
			\begin{table*}
				\caption{Spectral-type comparison for peculiar stars.}
				\label{tab:peculiars}
				\setlength{\tabcolsep}{14pt}
				\centering
				\begin{tabular}{lcc}
					\hline\hline
					LAMOST ID 			& Human type 			& \texttt{\textsc{NutMaat}}\\
					\hline
					J012145.51+443625.8	& kA1 hA5 V SrEu(Cr)	& A1 IV-V  SrSiEu\\
					J040517.81+335007.7	& kA0 hA5 V EuSrCr		& A1 II-III  Si\\
					J050755.19+495727.1	& A0 III SiSrCrEu		& A0 II-III  SrSiEu\\
					J051331.05+393032.4	& kA1 hA7 V SrEuCr		& A1 III-IV  SrSiEu\\
					J052856.69+475711.7	& kA1 hA7 V SrCrEu		& A1 IV-V  SrSiEu\\
					J053725.00+382012.2	& A0 III SiEuSrCr		& kA0 hA2 mA3  SiEu\\
					J054009.21+120945.0	& kA2 hA3 V SrCrEu		& A1 IV  SrSiEu\\
					J055527.76+152315.3	& kA2 hA5 V SrCrEu		& kA3 hA7 mA7  SiEu\\
					J065551.61+174248.1	& kA1 hA7 V SrCrEu		& kA1 hA5 mF0  SiEu\\
					J182441.87+485404.6	& kA0: hA5: V SrCrEu	& A0 III-IV  SrSiEu\\
					J190408.38+445439.1	& kA0 hA6 V EuCrSr(Si:)	& B8 V\\
					J190831.72+511959.7	& kA0 hA3 V EuSrCr(Si:)	& kA0 hA2 mA5  SiEu\\
					J194356.39+470831.0	& kA1 hA5 V SrEu(Cr)	& A0 IV-V  SrSiEu\\
					J212905.49+142755.4	& kA1 hA7 V SrCrEu(Si:)	& A1 IV  SrSiEu\\
					J213347.82+451647.6	& kA2 hA7 V SrCrEu		& A1 IV  SrSiEu\\
					\hline
				\end{tabular}
				\tablecomments{For clarity, the types are expressed in MK standard notation in which prefixes such as ``k'', ``h'', and ``m'' refer to spectral types inferred from the Ca$\,$\textsc{ii} K line, hydrogen lines, and metallic-line spectrum, respectively. For example, ``kA1 hA5 mF2'' indicates that the Ca$\,$\textsc{ii} line resembles that of an A1-type star, hydrogen lines match A5, and metal lines F2. Elemental peculiarities---such as Sr, Eu, or Cr---are appended at the end.}
			\end{table*}

			Currently, \texttt{\textsc{NutMaat}} supports the same classes of peculiarity handled by \texttt{MKCLASS}, including Am, Ap, $\lambda$ Boo, and composite types. It does not yet perform full classification of other types. However, \texttt{\textsc{NutMaat}} is capable of identifying emission-line spectra by detecting anomalous line profiles even if an exact corresponding type cannot yet be assigned. Extensions to the wavelength coverage and classification logic are planned to support these types explicitly in future releases (see Section \ref{sec:discussion}).
				
			To better diagnose the scope and limitations of \texttt{\textsc{NutMaat}}'s peculiarity detection, a larger sample of chemically peculiar stars---consistently classified and observed using a common instrument---will be curated in future work. This expanded set will allow for more statistically robust evaluation of the tool's performance across a wider range of spectral types, peculiarity classes, and data conditions. The current test, while limited in size, illustrates that \texttt{\textsc{NutMaat}} can serve as a reliable first-pass flagging tool for identifying spectra that merit expert follow-up, particularly among A- and early F-type stars where chemical anomalies are most common. In total, these results show that \texttt{\textsc{NutMaat}}'s handling of peculiarities is functional and interpretable, though still at an early stage of development. The current detection routines are intentionally conservative and do not yet address rare or more complex phenomena.
		
		\subsection{Classification Accuracy and Signal-to-noise} \label{subsec:s2n}
			To assess \texttt{\textsc{NutMaat}}'s robustness to decreasing S/N, we selected seven MK standard spectra from the \texttt{libnor36} library spanning the main full spectral sequence from O to M, and covering luminosity classes I, III, and V. Each spectrum was degraded to S/N levels of 5, 10, 20, 50, 100, 300, and evaluated five times at each level using different random noise seeds. Table \ref{tab:s2n} summarizes the most frequent spectral type and match quality output for each condition. At high S/N ($\ge 100$), \texttt{\textsc{NutMaat}}'s classifications are consistent and accurate across all types (except the O-type). At lower S/N levels, classification stability varies by type: early-type stars (O and B) exhibit moderate spectral drift under noise, while later types retain relatively robust classifications due to the presence of broader spectral features. In cases of tied classifications across replicates, a representative range is shown to reflect uncertainty. This shows that \texttt{\textsc{NutMaat}}'s tolerance to noisy spectra is comparable, if not superior, to the human ability for S/N $\le$ 20 \citep{Gray2014}.
			
			\begin{table*}
				\caption{\texttt{\textsc{NutMaat}} classification for seven MK standard stars under decreasing signal-to-noise ratios. Each entry reports the most frequently predicted spectral type and match quality label across five independent noise realizations. If no single type dominates, a range of classifications is shown.}
				\label{tab:s2n}
				\setlength{\tabcolsep}{12pt}
				\centering
				\begin{tabular}{clccclcc}
					\hline\hline
					True type 	& S/N		& Predicted type	& Quality	& 	True type 	& S/N		& Predicted type	& Quality	\\
					\hline
					O8 III 		& $\infty$	& O8 III			& Excel		& G5 Ib  		& $\infty$ 	& G5 Ib         	& Good 		\\
								& 300		& O7 V				& Excel		& 				& 300      	& G5 Ib         	& Good		\\
								& 100		& O6 II				& Vgood		&				& 100      	& G5 Ib         	& Good		\\
								& 50		& O6 II				& Vgood		&				& 50       	& G5 Ib         	& Good		\\
								& 20		& O9 III			& Good 		&				& 20       	& G5 Ib         	& Good		\\
								& 10		& O9 III			& Fair		&				& 10       	& G5--G9 Iab--III   & Good		\\
								& 5			& O6--G0 V			& Fair		&				& 5        	& G5--K0 Ia--III    & Fair		\\
					\hline
					B5 III 		& $\infty$ 	& B5 III--IV    	& Vgood 	& K5 III 		& $\infty$ 	& K5 II--III    	& Vgood 	\\
								& 300      	& B5 III--IV        & Vgood 	&				& 300      	& K5 II--III        & Vgood 	\\
								& 100      	& B5 III--IV        & Vgood 	&				& 100      	& K5 II--III        & Vgood 	\\
								& 50       	& B5 III--IV        & Vgood 	&				& 50       	& K5 II--III        & Vgood 	\\
								& 20       	& B5 III--IV        & Good 		&				& 20       	& K5 II--III        & Vgood 	\\
								& 10       	& B5 III--IV        & Good 		&				& 10       	& K5 III        	& Good 		\\
								& 5        	& B7--F4 IV--V      & Fair 		&				& 5        	& K0--K5 Iab--V     & Fair 		\\
					\hline
					A0 V   		& $\infty$ 	& A0 V          	& Excel 	& M2 III 		& $\infty$ 	& M2 III        	& Vgood 	\\
								& 300      	& A0 V          	& Excel 	&				& 300      	& M2 III        	& Vgood 	\\
								& 100      	& A0 V          	& Vgood 	&				& 100      	& M2 III        	& Vgood 	\\
								& 50       	& A0 V          	& Vgood 	&				& 50       	& M2 III        	& Vgood 	\\
								& 20       	& A0 V          	& Good 		&				& 20       	& M2 III       		& Vgood 	\\
								& 10       	& A0 V          	& Good 		&				& 10       	& M1--M2 II--IV     & Vgood 	\\
								& 5        	& A3--F4 V      	& Fair 		&				& 5        	& M0--M2 II--IV    	& Good 		\\
					\hline
					F5 Ib  		& $\infty$ 	& F5 Ib         	& Excel 	&				&			&					&			\\
								& 300      	& F5 Ib--II         & Excel 	&				&			&					&			\\
								& 100      	& F5 Ib--II         & Excel 	&				&			&					&			\\
								& 50       	& F5 Ib--II     	& Vgood 	&				&			&					&			\\
								& 20       	& F4--F6 Iab--III   & Good 		&				&			&					&			\\
								& 10       	& F3--F6 Ia--IV     & Good 		&				&			&					&			\\
								& 5        	& F4--F8 II--V     	& Fair 		&				&			&					&			\\
					\hline
				\end{tabular}
			\end{table*}
	
	\section{The MaStar Catalog}	\label{sec:nm-mastar}
		To demonstrate the applicability of \texttt{\textsc{NutMaat}} to modern, large-scale spectroscopic datasets, we applied the tool to the spectra of the MaStar library \citep{Yan2019} from SDSS DR17 \citep{Abdurrouf2022}. MaStar is a comprehensive empirical stellar library constructed using the same BOSS spectrographs \citep{Smee2013} as the MaNGA galaxy survey \citep{Bundy2014}, with coverage over the 3622--10354 \r{A} wavelength range at R $\sim$ 1800. The targets in the MaStar library were selected to span a wide range of atmospheric parameter space. Unlike other libraries used in Section \ref{sec:tests}, the MaStar pipeline does not derive or publish MK spectral classifications. While spectral types may appear in some catalog products, these are based on cross-matches with external sources via SIMBAD and are not consistently generated. Instead, MaStar provides only stellar atmospheric parameters---effective temperature ($T_{\text{eff}}$), surface gravity ($\log{g}$), and metallicity ([Fe/H]). As such, the classifications generated by \texttt{\textsc{NutMaat}} constitute a value-added product, providing uniform MK types and luminosity classes for the included spectra. This offers users an alternative labeling scheme, particularly useful for empirical template matching and population synthesis modeling, where MK-based consistency is often preferred over parameter-based templates (e.g., \citealp{Bruzual2003}; \citealp{Conroy2013}).
			
		We specifically used MaStar visit spectra flagged as high-quality\footnote{\url{https://data.sdss.org/sas/dr17/manga/spectro/mastar/v3_1_1/v1_7_7/mastar-goodspec-v3_1_1-v1_7_7.fits.gz}}---those without visual reduction issues, poor calibration, or low S/N---as documented in \citet{Yan2019}. This choice was made to ensure a stable foundation for classification and to avoid conflating algorithmic performance with data quality issues. Future work may expand this analysis to include lower-quality spectra to further test \texttt{\textsc{NutMaat}}'s robustness under less ideal observational conditions.
			
		The goal here was not to test \texttt{\textsc{NutMaat}}, but to generate a value-added catalog of MK classifications for a widely-used, high-fidelity stellar library. In this way, \texttt{\textsc{NutMaat}} provides an additional layer that complements the atmospheric parameters derived by other tools. A sample of this catalog is shown in Table \ref{tab:mastar}. The full MK classification catalog for MaStar spectra is publicly released alongside this manuscript. For each spectrum, \texttt{\textsc{NutMaat}} returns a raw $\chi^2$ value quantifying the deviation between the input and the best-matching MK standard from the chosen library (\texttt{libnor36} in this case), computed over the classification window (3800--5600 \r{A}). While the software does not currently compute reduced $\chi^2$ internally, users may interpret the reported values in light of the approximate degrees of freedom involved. Since classification typically uses $\sim 1750$ valid wavelength bins and involves 3--5 effective fitted parameters (e.g., for interpolation, radial velocity shift, and flux normalization), we estimate the degrees of freedom for this set to be $\nu = 1700 \pm 50$. Thus, an approximate reduced $\chi^2$ may be obtained as $\chi^2/1700$, enabling standard comparison across spectra. Although Table \ref{tab:mastar} lists raw $\chi^2$ values, this scaling factor offers a consistent interpretive framework. Future versions of \texttt{\textsc{NutMaat}} will include explicit output of reduced $\chi^2$ based on the choice of spectrum and standard library.
			
		\texttt{\textsc{NutMaat}} also returns a categorical \emph{match quality flag} for each classification to indicate the confidence level of the assigned MK type. This flag is derived from the $\chi^2$ value of the best-match spectrum. The threshold values used to determine the flag vary slightly with spectral type, reflecting the differing line strengths and classification complexity across the HR diagram. For late-type stars (late K to M), higher $\chi^2$ values are tolerated than for hotter stars. The flags take one of five values: `Excel', `Vgood', `Good', `Fair', or `Poor', corresponding to increasingly weaker matches between the input spectrum and the MK standards. The thresholds are empirically calibrated; for example, $\chi^2$ values below $10^{-4}$ are always rated `Excel', while values above $10^{-1}$ (late types) or $5\times10^{-2}$ (early types) are flagged as `Poor'. These qualitative labels enable users to filter or weigh classifications based on match quality for further analysis. In addition, \texttt{\textsc{NutMaat}} classifications may be cross-referenced against MaStar parameters\footnote{\url{https://data.sdss.org/sas/dr17/manga/spectro/mastar/v3_1_1/v1_7_7/vac/parameters/v2/mastar-goodstars-v3_1_1-v1_7_7-params-v2.fits}} via the included MaNGA IDs and coordinates.
		
		\begin{table*}
			\caption{A sample of the MaStar library classification catalog using \texttt{\textsc{NutMaat}}. For each star, we list the MaNGA ID, equatorial coordinates, signal-to-noise ratio, \texttt{\textsc{NutMaat}} spectral type and luminosity class, match quality, and $\chi^2$ fit statistic. The $\chi^2$ values listed are raw ones returned by \texttt{\textsc{NutMaat}}. For interpretive purposes, the degrees of freedom are estimated to be $\sim 1700$. An approximate reduced $\chi^2$ can thus be calculated as $\chi^2/1700$.}
			\label{tab:mastar}
			\setlength{\tabcolsep}{14pt}
			\centering
			\begin{tabular}{lcccccc}
				\hline\hline
				MaNGA ID	& $\alpha$ (Degree) & $\delta$ (Degree) & S/N 		& Spectral type		& Quality 	& $\chi^2$\\
				\hline
				5-12626		& 135.58530			& 57.57387			& 54		& G0 IV			& Good		& $3.45 \times 10^{-3}$\\
				5-66039     & 134.32357			& 57.87625			& 53		& A4 Iab			& Good   	& $1.56 \times 10^{-3}$\\
				5-108715    & 132.75460			& 57.95367			& 57		& F9 IV-V		& Vgood  	& $6.19 \times 10^{-4}$\\
				7-17443125  & 290.26975			& 51.34533			& 228		& K3 III   		& Good   	& $2.33 \times 10^{-3}$\\
				7-18131529  & 296.92869			& 43.79183			& 49		& B3 II-III		& Vgood  	& $9.29 \times 10^{-4}$\\
				\hline
			\end{tabular}
			\tablecomments{Table \ref{tab:mastar} is published in its entirety in the machine-readable format. A portion is shown here for guidance regarding its form and content.}
		\end{table*}
	
	\section{Limitations and Future Work}	\label{sec:discussion}
		While \texttt{\textsc{NutMaat}} demonstrates promising performance in classifying stellar spectra across a wide range of spectral and luminosity classes, several limitations---most of which directly inherited from \texttt{MKCLASS}---remain that will be addressed in future development cycles. First, the current implementation is constrained to the 3800--5600$\,$\r{A} wavelength range, which excludes important diagnostic features used to distinguish later M types, identify emission-line objects, and refine luminosity subclasses in cooler stars. This limitation contributes to certain degeneracies observed in the confusion matrix (see Figure \ref{fig:confusion-matrix}) and hinders robust identification of some classes. Second, \texttt{\textsc{NutMaat}} currently models only a limited set of spectral peculiarities---specifically those implemented within \texttt{MKCLASS}, such as Am, Ap , $\lambda$ Boo, and metal-poor stars. Other peculiar classes are not yet systematically handled. Although some emission-line spectra are flagged as anomalous, their exact-type classification remains beyond \texttt{\textsc{NutMaat}}'s present capabilities. Extending peculiarity logic to include a broader set of features and anomaly detection routines is an ongoing priority. Similarly, the subroutines for classifying O-type stars are currently rudimentary due to the insufficient number of O-type spectra available for testing. Third, luminosity classification accuracy remains limited in part by the coarseness of the calibration datasets available, especially in empirical libraries with sparse coverage of early- and late-type giants and supergiants. Additional benchmarking across datasets with finely graded luminosity standards is needed to refine this component of the classifier. In addition, \texttt{\textsc{NutMaat}} currently uses \texttt{SciPy} \citep{Virtanen2020} modules to perform the numerical operations necessary for preprocessing and classification, including continuum fitting, spectral interpolation, line and feature localization, and both uni- and multivariate function optimization. These routines replicate the internal logic of \texttt{MKCLASS}, such as spectral index computation, line profile analysis, and feature reconciliation across modules, which are described in detail in \citet{Gray2014} and the \texttt{MKCLASS} documentation \citep{Gray2014a}. For example, while \texttt{MKCLASS} uses routines like \texttt{ROUGHTYPE1} and \texttt{ROUGHTYPE2} to initiate classification based on $\chi^2$ fitting and index-based templates, \texttt{\textsc{NutMaat}} implements analogous logic using Python's \texttt{scipy.interpolate} and \texttt{scipy.optimize} modules, allowing template matching and feature alignment to be modular and reproducible. Although this Pythonic reimplementation prioritizes transparency and adaptability, it comes at a computational cost.
		
		To address these limitations, planned development efforts will focus on the following:
		\begin{enumerate}
			\item Expanding the spectral coverage beyond 5600 \r{A} to include redder Balmer lines and molecular bands critical for M-type stars and emission-line detection.
			\item Developing classification logic for a broader range of stellar peculiarities and O-type stars by incorporating additional spectral features for classes such as carbon stars, white dwarfs, Wolf-Rayet stars, etc. 
			\item Curating an expanded benchmark set of chemically peculiar stars drawn from uniformly classified and consistently observed spectra, to rigorously evaluate \texttt{\textsc{NutMaat}}'s peculiarity recognition performance across a wider domain of the HR diagram.
			\item Developing noise-aware optimization routines to minimize divergence artifacts observed in a minority of outliers.
			\item Speeding up the program by adopting \texttt{pandas} data frames for standard libraries integration to save some of the overhead cost, and by using an approach like the Cython language \citep{Dalcin2011}.
			\item Improving the user interface and supporting documentation by developing a formal user guide, as well as a Wiki page on the official \texttt{\textsc{NutMaat}} GitHub repository. These materials will include detailed usage instructions, interpretive guidance for the output flags, and illustrative examples to facilitate adoption by the broader community.
		\end{enumerate}
		
		As noted in Section \ref{sec:nutmaat}, \texttt{\textsc{NutMaat}} will eventually be evaluated in parallel with other publicly available automated classifiers, including \texttt{MKCLASS}, \texttt{\textsc{PyHammer}}, and \texttt{STARMIND}. Each classification tool will be evaluated using a consistent set of metrics, including classification accuracy, systematic offsets, and computational efficiency. Particular attention will be paid to the tools' abilities to handle chemically peculiar stars, where we will test classification robustness using known peculiar spectra. We also aim to evaluate not just scientific performance, but also workflow efficiency, measuring the time, input complexity, and preprocessing requirements of each tool. All code, data, and evaluation metrics will be made publicly available to ensure reproducibility. These directions reflect our commitment to making \texttt{\textsc{NutMaat}} a reliable, extensible, and scientifically valuable tool for stellar classification in both archival and future spectroscopic surveys.
	
	\section{Conclusions}	\label{sec:conclusions}
		In this paper, we presented \texttt{\textsc{NutMaat}}, a Python-based stellar spectral classification package developed to replicate the functionality of \texttt{MKCLASS} while enhancing usability and scalability by using up-to-date computational tools. We adopted an object-oriented framework and integrated flexible data structures to increase the efficiency of batch processing. \texttt{\textsc{NutMaat}} offers platform independence and seamless integration into astronomical survey pipelines.
		
		We tested \texttt{\textsc{NutMaat}} on benchmark datasets (CFLIB and MILES) to demonstrate its capabilities in spectral and luminosity classification. The results aligned closely with the performance of \texttt{MKCLASS}, with \texttt{\textsc{NutMaat}} slightly favoring later spectral types. The package also exhibited tolerance to low S/N, reliably classifying spectra down to S/N $\approx$ 10, comparable to human capabilities at S/N $\le$ 20.
		
		We applied \texttt{\textsc{NutMaat}} to the MaStar library from SDSS DR17 to further demonstrate its practicality for large scale surveys, enabling the release of a diverse stellar classification catalog that can be used for empirical template matching and population synthesis. This illustrates \texttt{\textsc{NutMaat}}'s ability not only to replicate expert classification but to enhance scientific utility of archival datasets that do not include MK types generated by a uniform pipeline. Nevertheless, limitations inherited from \texttt{MKCLASS} persist, including rudimentary classification procedures for O-type and non-canonical stars, and reliance on the 3800-5600 \r{A} spectral range; in addition to slower execution due to Python's interpretative nature.
		
		Future development of \texttt{\textsc{NutMaat}} will prioritize extending spectral criteria to a broader wavelength range, refining the classification procedures for peculiar spectra, optimizing computational efficiency via Cython integration, and enhancing MK-standards library customization using \texttt{pandas} data structures. This will enable \texttt{\textsc{NutMaat}} to maintain the fidelity of the MK system while adapting to modern computational workflows, curating for the demands of big-data astronomy and next-generation spectroscopic surveys. \texttt{\textsc{NutMaat}} is publicly available and open-source, with a user guide and documentation in preparation to support broader community use.
		
	\begin{acknowledgements}
		\software{
			Python \citep{van1995python},
			\texttt{NumPy} \citep{Harris2020},
			\texttt{pandas} \citep{pandas2024,McKinney2010},
			\texttt{Matplotlib} \citep{Hunter2007},
			\texttt{Astropy} \citep{Robitaille2013,PriceWhelan2018,Collaboration2022},
			\texttt{SciPy} \citep{Virtanen2020}
		}
		
		We gratefully acknowledge the foundational contributions of the \texttt{MKCLASS} software \citep{Gray2014}, which inspired and informed the development of \texttt{\textsc{NutMaat}}. The methodologies, spectral criteria, and expert-system logic pioneered by \texttt{MKCLASS} remain integral to \texttt{\textsc{NutMaat}}’s design.
		
		This research has made use of the VizieR catalogue access tool, CDS,
		Strasbourg, France (DOI: 10.26093/cds/vizier). The original description 
		of the VizieR service was published in \citet{Ochsenbein2000}.
		
		This work has used the LAMOST database. Guoshoujing Telescope (the Large Sky Area Multi-Object Fiber Spectroscopic Telescope LAMOST) is a National Major Scientific Project built by the Chinese Academy of Sciences. Funding for the project has been provided by the National Development and Reform Commission. LAMOST is operated and managed by the National Astronomical Observatories, Chinese Academy of Sciences.
		
		SDSS-IV acknowledges support and resources from the Center for High Performance Computing  at the University of Utah. The SDSS website is \url{www.sdss4.org}.
		
		SDSS-IV is managed by the Astrophysical Research Consortium for the Participating Institutions of the SDSS Collaboration including the Brazilian Participation Group, the Carnegie Institution for Science, Carnegie Mellon University, Center for Astrophysics | Harvard \& Smithsonian, the Chilean Participation Group, the French Participation Group, Instituto de Astrof\'isica de Canarias, The Johns Hopkins University, Kavli Institute for the Physics and Mathematics of the Universe (IPMU) / University of Tokyo, the Korean Participation Group, Lawrence Berkeley National Laboratory, Leibniz Institut f\"ur Astrophysik Potsdam (AIP),  Max-Planck-Institut f\"ur Astronomie (MPIA Heidelberg), Max-Planck-Institut f\"ur Astrophysik (MPA Garching), Max-Planck-Institut f\"ur Extraterrestrische Physik (MPE), National Astronomical Observatories of China, New Mexico State University, New York University, University of Notre Dame, Observat\'ario Nacional / MCTI, The Ohio State University, Pennsylvania State University, Shanghai Astronomical Observatory, United Kingdom Participation Group, Universidad Nacional Aut\'onoma de M\'exico, University of Arizona, University of Colorado Boulder, University of Oxford, University of Portsmouth, University of Utah, University of Virginia, University of Washington, University of Wisconsin, Vanderbilt University, and Yale University.
	\end{acknowledgements}
	
	\begin{acknowledgements}
		We are grateful to the anonymous referee for their insightful comments and suggestions, which have significantly improved the quality and clarity of this manuscript.
	\end{acknowledgements}
	
	\bibliography{refs}{}
	\bibliographystyle{aasjournal}
		
\end{document}